\documentclass[a4paper,10pt]{article}

\usepackage{textcomp}
\usepackage{amsmath}
\usepackage{amssymb}
\usepackage{latexsym}
\usepackage{color}

\begin{document} 

\newcommand{\junk}[1]{}
\newtheorem{theorem}{Theorem}
\newtheorem{corollary}{Corollary}
\newtheorem{lemma}{Lemma}
\newtheorem{conjecture}{Conjecture}
\newtheorem{claim}{Claim}
\newtheorem{fact}{Fact}
\newtheorem{remark}{Remark}

\newtheorem{definition}{Definition}
\newtheorem{assumption}{Assumption}
\newtheorem{observation}{Observation}
\newtheorem{proposition}{Proposition}
\newtheorem{example}{Example}
\newtheorem{open}{Open Problem} 

\newcommand{\qed}{\rule{7pt}{7pt}}
\newcommand{\order}[1]{O(#1)}
\newenvironment{proof}{\noindent\textbf{Proof:}}{\qed\bigskip}
\newenvironment{proof_sketch}{\noindent\textbf{Sketch of
Proof}\hspace*{1em}}{\qed\bigskip}
\newenvironment{proof_of_claim}{\noindent\textbf{Proof of
Claim:}}{\qed\bigskip}

\newcommand{\smallspacing}{\baselineskip = 0.98\normalbaselineskip}
\newcommand{\tinyspacing}{\baselineskip = 0.7\normalbaselineskip}
\newcommand{\morespacing}{\baselineskip = 1.1\normalbaselineskip}
\newcommand{\normalspacing}{\baselineskip=.985\normalbaselineskip}

\def\floor#1{\ensuremath{\left\lfloor{#1}\right\rfloor}}
\def\root#1{\ensuremath{\sqrt{\smash[b]{#1}}}}
\def\Rmin{\ensuremath{r_{\text{min}}}}
\def\Rmax{\ensuremath{r_{\text{max}}}}
\def\OPT{\textrm{OPT}}
\def\dotminus{\stackrel{.}{-}}
\def\OR{\vee}
\def\AND{\wedge}
\def\bigdoublewedge{\bigwedge}
\def\bigdoublevee{\bigvee}
\def\goesto{\rightarrow}

\def\beginproof{\noindent{\bf Proof.}\quad}
\def\endproof{\qed}

\title{Some Remarks on Almost Periodic \\ Sequences and Languages}

\author{GABRIEL ISTRATE \\ Faculty of Mathematics, University of Bucharest \\ Str. Academiei 14, Bucharest, Romania}
\date{}
\maketitle

\section{Introduction}

Almost periodicity has been considered in Formal Language Theory in connection with some topics in Symbolic Dynamics (see e.g. \cite{morse1938symbolic}). A notorious example of 
an almost-periodic sequence is the famous Thue-Morse sequence (\cite{jacobs1969maschinenerzeugte}, \cite{marcus2007infinite}). Almost periodicity has been considered when dealing with the decidability of monadic theories of unary functions \cite{semenov1984decidability}. In \cite{marcus2007infinite} some problems concerning this property are raised. For instance, one asks whether or not there exists some almost periodic word $\alpha$ such that $Sub(\alpha)$, the language of its finite factors, is context-free non-regular. 

We will answer negatively (even in a stronger form) this question, as well as discussing other related topics. 

\section{Results} 

We will use Formal Language Notations from \cite{paun2013marcus}, \cite{salomaa1987formal}. By $V^{\omega}$ we will denote the set of one-sided infinite words over the alphabet $V$. $V^{\omega^{*}+\omega}$ will mean its two-sided counterpart. For an infinite $\alpha$ we denote by $Sub(\alpha)$ the language of its finite factors. $z<w$ will mean that $z$ is a factor of $w$. Note that the meaninig of almost-periodicity from \cite{semenov1984decidability} is less restrictive than the one in \cite{marcus2007infinite}. It is this latter version we will deal with: 

\begin{definition} A sequence $\alpha \in V^{\omega}$ is called \emph{almost periodic} iff for any $w<\alpha$ there exists a positive integer $n_w$ such that any factor $z$ of $\alpha$ of length at least $n_w$ has $w$ as a factor. 
\end{definition} 

Let us move this definition of almost-periodicity from infinite sequences to languages: 

\begin{definition} 
A language $L\subset V^{*}$ is called 
\begin{itemize} 
\item \emph{closed} iff $L$ is closed under taking subwords. 
\item \emph{confluent} iff for any $x$ and $y$ in $L$ we may find $z\in L$ such that $x<z$ and $y<z$. 
\item \emph{redundant} iff for any $x\in L$ we may find $n_x\geq 1$ such that for any $z\in L$, $|z|\geq n_x$ implied $x<z$. 
\item \emph{almost periodic} iff $L$ is closed and redundant. 
\end{itemize} 
\end{definition}

\noindent\textbf{Remarks}
\begin{itemize} 
\item[(a).] If $L$ is almost periodic then $L$ is confluent. 
\item[(b).] The following assertions are equivalent: 
\begin{itemize} 
\item[-] $L$ is infinite, closed and confluent. 
\item[-] $L=Sub(\alpha)$ for some $\alpha \in V^{\omega^{*}+\omega}$
\end{itemize} 
\item[(c).] The following assertions are equivalent: 
\begin{itemize} 
\item[-] $L$ is infinite and almost periodic.  
\item[-] $L=Sub(\alpha)$ for some almost periodic $\alpha \in V^{\omega^{*}+\omega}$. 
\end{itemize} 
\item[(d).] If $\alpha \in V^{\omega}$ is almost periodic then $Sub(\alpha)$ is almost periodic. 
\end{itemize} 
\textbf{Justification.} To prove the non-tricial part of (b), we enumerate $L$ as $x_0,x_1,\ldots,$ and define $y_n\in L$ by $y_0=x_0,y_n<y_{n+1}, x_{n}<x_{n+1}$ (by the confluence of $L$). It follows that $y_0<y_1<\ldots <y_n<\ldots$. Taking the (bilateral) limit we find $\alpha \in V^{\omega^{*}+\omega}$ with the desired properties. Now (c) follows by combining (a) and (b). $\Box$

\begin{definition} 
A family $\mathcal{F}$ of languages avoids almost periodicity iff any almost periodic language in $\mathcal{F}$ is regular. 
\end{definition} 

Now we may state: 

\begin{theorem} 
Let $\mathcal{F}$ be a family of languages such that for any infinite $L$ in $\mathcal{F}$ one may find $w\neq \lambda$ such that $\{w^{n}|n\geq 1\}\subset Sub(L)$. Then $\mathcal{F}$ avoids almost periodicity. 
\label{thm-one}
\end{theorem} 

\textit{Proof.} 
Take $L\in \mathcal{F}$ infinite and almost periodic and consider the corresponding $w$. As $lim_{n} |w|^n=\infty$ it follows (from the definition of almost periodicity) that $L\subset Sub(\{w^n|n\geq 1\})$. On the other hand $\{w^{n}|n\geq 1\}\subset Sub(L)$ implies $Sub(\{w^{n}|n\geq 1\})\subset Sub(Sub(L))=Sub(L)=L$, hence $L= Sub(\{w^n|n\geq 1\})$ is a regular language. $\Box$

\textbf{Remark.} Many families of languages, including $\mathcal{L}_{2}$ (the family of context-free languages), $\mathcal{M}_{f}$ (the family of matrix languages of finite index \cite{dassow1989regulated}), $\mathcal{SM}_{2}$  (the family of simple matrix languages \cite{dassow1989regulated}), $\mathcal{C},\mathcal{G}$ (the families of external contextual and of generalized contextual languages \cite{marcus69contextual}), $\mathcal{I}, \mathcal{IS}$ (the families of internal contextual and of internal contextual with choice languages \cite{paun2013marcus}) satisfy the requirements of our lemma. 

\begin{corollary} 
If $L=Sub(\alpha)$ for some almost periodic $\alpha \in V^{\omega}$ (or $V^{\omega^{*}+\omega}$) and $L$ belongs to one of the above mentioned families of languages then $L$ is regular. 
\end{corollary} 

Another open problem from \cite{marcus2007infinite} was the existence of an algorithm deciding whether a given regular language can be written as $L=Sub(\alpha)$ for some $\alpha \in V^{\omega}$. We still cannot answer this question. However, by restricting ourselves to almost periodic sequences we get a better situation: 

\begin{corollary} Let $\mathcal{F}$ be a family of languages having the following properties: 
\begin{itemize} 
\item The finiteness problem for $\mathcal{F}$ is decidable. 
\item $\mathcal{F}$ constructively satisfies the (proof of) Theorem~\ref{thm-one} (i.e. there is an algorithm which, given $L\in \mathcal{F}$ infinite, finds the right $w$ and then tests whether $L=Sub(\{w^{n}|n\geq 1\})$. 
\end{itemize} 
Then it is decidable whether $L\in \mathcal{F}$ can be written as $L=Sub(\alpha)$ for some almost periodic $\alpha\in V^{\omega}$. 
\label{cor-2}
\end{corollary} 
\textbf{Remark.} A sufficient condition for the validity of the second condition in the previous corollary is the following: 
\begin{itemize} 
\item one can effectively construct the required $w$; 
\item given $L_1\in \mathcal{F}$ and $L_2\in \mathcal{L}_{3}$ one can effectively check whether $L_1=L_2$. 
\end{itemize} 
\begin{corollary} The following families satisfy the hypothesis of Corollary~\ref{cor-2} (and hence the problem whether an arbitrary language in them can be written as $L=Sub(\alpha)$ for $\alpha\in V^{\omega}$ almost periodic is decidable): 
\begin{itemize} 
\item the family of regular languages. 
\item the family of unambiguous context-free languages. 
\end{itemize} 
\label{cor-3} 
\end{corollary}
\textit{Proof.} Both these families of languages satisfy the conditions of the previous remark (for unambiguous context-free languages one uses a result due to Semenov, see  \cite{beauquier1985rational,salomaa90fps}) $\Box$ 

\begin{open} 
Are Semenov-type results true for the families $\mathcal{M}_{f}$, $\mathcal{SM}_{2}$ ? 
\end{open} 
\begin{open} What is the decidability status of this problem for families $\mathcal{M}_{f}$, $\mathcal{SM}_{2}$, $\mathcal{IS}$ ? 
\end{open} 

We could add the families $\mathcal{C},\mathcal{G},\mathcal{I}$ to the list from Corollary~\ref{cor-3} . However, for these families we may state a more precise result. 

\begin{theorem} Let $L$ be an infinite almost-periodic language in one of the families $\mathcal{C},\mathcal{G},\mathcal{I}$. Then we can find $a\in V$ such that $L=a^{*}$.
\end{theorem} 
\textit{Proof.} We will prove our result for the family $\mathcal{I}$ (the other cases are analogous). Let $L=L(V,B,C)$ be an infinite almost-periodic language in $\mathcal{I}$ such that $L=\{a\}^{*}$ for no $a$ in $V$. Then $L$ must include at least two different letters $a$ and $b$ (for the only infinite almost-periodic language over a one letter alphabet a is $\{a\}^{*}$). Consider $x$, a nonvoid candidate for $w$ (i.e. $\{x^{n}|n\geq 1\}\subset Sub(L)$). It follows that $|x|\geq 2$. Indeed, suppose that $x\in V$. As 
$\{x^{n}|n\geq 1\}\subset Sub(L)$ (from the definition of family $\mathcal{I}$) it would follow that $L=\{x\}^{*}$, which is not the case. As $L$ is closed under taking subwords, $a,b\in L$. It follows that $a,b\in \mathcal{B}$ (any context must increase the length by at least two: this follows the same way as $|x|\geq 2$). Take $w\in V^{*}$, $w\neq \lambda$ having minimal length such that $\{aw^{n}|n\geq 1\}\cup \{bw^{n}|n\geq 1\}\subset Sub(L)$ or $\{w^{n}a|n\geq 1\}\cup \{w^{n}b|n\geq 1\}\subset Sub(L)$. Clearly there exists such a $w$ (any nonvoid semicontext is a candidate for $w$). Suppose we are in the first case and, moreover, $w$ does not end in $a$ (if not then exchange $a$ and $b$). Clearly $|w|\geq 2$. As $w^n\in Sub(L)$ for any $n$ and $L$ is almost-periodic, from the proof of Theorem~\ref{thm-one} it follows that $L=Sub(\{w^{n};n\geq 1\})$ hence $aw\in Sub(L)=L=Sub(\{w^{n};n\geq 1\})$. 

As $|w|\geq 2$ and $a$ is not the last letter in $w$, $aw<w^2$, hence $w^2=zawt$ with $za\in Pref(w)$, $t\in Suff(w)$. As $|za|+|t|=|w^2|-|w|=|w|$, it follows that $w=zat$, hence $zazatt=zatzat$ so $(za)t=t(za)$. The equation $uv=vu$ has as solutions the system $\{u=\beta^{m}, v=\beta^{n}|\beta\in V^{*},m,n\geq 1\}$. As $t\neq \lambda$ (otherwise $\alpha$ would have been the last letter of $w$) we have $w=(za)t=\beta^{k}$ for some $\beta\in V^{*}, \beta\neq \lambda$ and $k\geq 2$. 
But then $\{a\beta^{n}|n\geq 1\}\cup \{b\beta^{n}|\beta\geq 1\}\subset Sub(L)$ and $|\beta|<|w|$, contradicting the minimality of $w$. Hence $L=\{a\}^{*}$. $\Box$

Let us return to the problem of testing whether a given regular language $L$ can be written as $L=Sub(\alpha)$ for some $\alpha\in V^{\omega}$. The bi-sided version of this problem seems easier to tackle. Indeed, $L=Sub(\alpha)$ for some $\alpha\in V^{\omega^{*}+\omega}$ iff $L$ is infinite, closed and confluent. Finiteness and closure are decidable for regular languages: given a regular grammar $G_1$, construct (effectively) a grammar $G_2$ for $Sub(L(G_1))$ and then test whether $L(G_1)=L(G_2)$. Let us further note that $L$ is confluent iff $Sub(L)$ is confluent. Now it is straightforward that testing whether a given regular language can be written in the required form is algorithmically equivalent to the problem of testing confluence for regular languages. 

\begin{open} 
Is confluence open for regular languages?
\end{open} 

\textbf{Note (2022):} The Open Problem 3 has been solved (affirmatively) in Harju, Tero, and Lucian Ilie. "On quasi orders of words and the confluence property." Theoretical Computer Science 200.1-2 (1998): 205-224.

\bibliographystyle{abbrv}
\bibliography{/Users/gistrate/Dropbox/texmf/bibtex/bib/bibtheory}
\end{document}